# Theoretical study on the stimulated Brillouin scattering in a sub-wavelength anisotropic waveguide: Acousto-optical coupling coefficients and effects of transverse anisotropies

Xiao-Xing Su[1,2], Xiao-Shuang Li[1], Yue-Sheng Wang[1] and Heow Pueh Lee[2]

[1]Institute of Engineering Mechanics, Beijing Jiaotong University, Beijing 100044, China
[2]Department of Mechanical Engineering, National University of Singapore, Singapore 117576, Singapore

**Abstract:** A theoretical study on the stimulated Brillouin scattering (SBS) in a sub-wavelength anisotropic waveguide is conducted. The optical, photoelastic and mechanical anisotropies of the waveguide materials are all taken into account. First, the integral formulae for calculating the acousto-optical coupling coefficients (AOCCs) due to the photoelastic and moving interface effects in SBS are extended to an optically anisotropic waveguide. Then, with the extended formulae, the SBSs in an elliptical nanowire with strong transverse anisotropies are simulated. In the simulations, the elastic fields are computed with the inclusion of mechanical anisotropy. Observable effects of the strong transverse anisotropies are found in numerical results. Most notably, the SBS gains of some elastic modes are found to be very sensitive to the small misalignment between the waveguide axes and the principal material axes. Detailed physical interpretations of this interesting phenomenon are provided. This interesting phenomenon implies an attractive way for more sensitive tuning of the SBS gain without significantly changing the phononic frequency.

## 1. Introduction

The stimulated Brillouin scattering (SBS) [1~3] is a three-order nonlinear optical process in which the pump, Stokes photons and the associated hypersonic phonons collectively interact. Like other nonlinear optical phenomena, the SBS is usually very significant in a waveguide (typically an optical fiber). It may induce undesirable noise, or be utilized to enable many practical applications including the Brillouin lasers or sensors [3, 4], slow or fast light [5, 6], efficient hypersonic phonon generation [7~10], optical pulse compression [11, 12] and so on. In bulk systems or waveguides at micron scales, the SBS is mainly mediated through the photoelastic (PE) responses of materials and, inversely, the optical forces due to electrostriction. With the rapid development of integrated optics and sub-wavelength/nano optics and also with the emergence of nano opto-mechanics [13~15], more and more attention has been paid to the SBS at sub-wavelength scales [16~44]. Owing to the large surface-to-volume ratio at sub-wavelength scales, the perturbations on the optical waves due to the motion of a sharp interface (namely, the moving interface (MI) effect or moving boundary effect [44~48]) as well as optical radiation pressures [49~51] may also become physically relevant coupling mechanisms of SBS.

The behaviors of SBS may change radically at sub-wavelength scales. Rakich et al. [16~18] demonstrated that the SBS could be strongly dependent on the waveguide geometry at sub-wavelength scales, and large optical radiation pressure on boundaries could induce giant enhancement of SBS gain even for the generally weak forward SBS (FSBS). Large SBS gain enhancements due to the new coupling mechanisms were also reported by other researchers [19~25]. Among them, the studies on the FSBS in a hybrid photonic-phononic crystal waveguide which was not longitudinally invariant were reported in Refs. [24, 25]. On the other hand, different coupling mechanisms may interfere destructively to decrease the SBS gain dramatically. Florez et al. [26] found experimentally a Brillouin self-cancellation phenomenon, which arose from exactly opposing acousto-optical coupling coefficients (AOCCs) due to the PE and MI effects. Besides the new coupling mechanisms, surface acoustic waves [27] or even leaky acoustic modes [28] were also found to get involved in SBS at sub-wavelength scales. These new findings had significant implications for the development of the state of the art SBS-active devices [29~32].

In this new field of SBS, some researchers [33~41] had established systematic and comprehensive theoretical models. Among them, Wolff et al. [36] presented a systematic coupled-mode formulation, in which detailed derivations of the integral formulae for the AOCCs due to all the known coupling mechanisms in SBS were given. Sipe et al. [37] recast the problem in a Hamiltonian based framework, also leading to the derivations of these AOCCs. These theoretical models provide mathematically rigorous and physically clear analysis of the coupling mechanisms in SBS. However, we notice that the existing integral formulae for AOCCs of SBS cannot take into account the anisotropy of optical permittivity. In fact, there are very few studies on SBS considering material anisotropies, whether at bulk, micron or nano scales. Wolff et al. [20] included in their theoretical design the mechanical and photoelastic anisotropies of the considered material Germanium which were found to cause significant effects on SBS. However, optical anisotropy was not addressed by them due to the optically isotropy of Germanium. Smith et al. [42, 43] studied the SBS in meta-materials and found large SBS enhancement. However, for the cubic-latticed meta-materials they considered, the effective optical permittivities were still isotropic in the long wavelength limit.

In the present work, a theoretical study on the SBS in a sub-wavelength anisotropic waveguide is conducted. The optical, photoelastic and mechanical anisotropies of the waveguide materials are all taken into account. First, the integral formulae for calculating AOCCs due to the PE and MI effects in SBS are extended to an optically anisotropic waveguide. Then, with the extended formulae, the SBSs in an elliptical nanowire with strong transverse anisotropies are simulated. In the simulations, the elastic fields are computed by including the mechanical anisotropy. Observable effects of the strong transverse anisotropies on SBS gains are found in numerical results.

## 2. Fundamentals of the work

### 2.1 Description of the physical fields involved in SBS

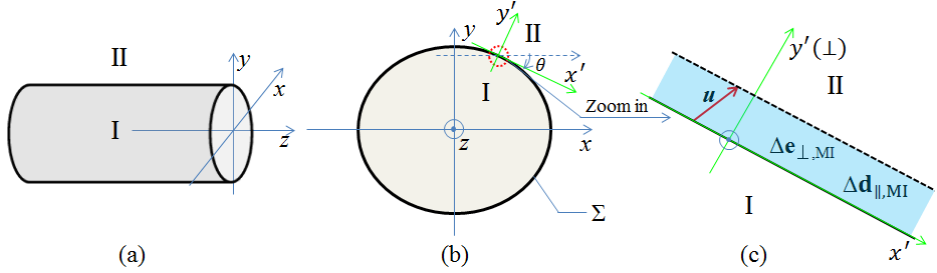

Fig. 1 (a) Schematic sketch of a waveguide which is translationally invariant in the $z$ direction. The core and cladding are filled with materials I and II, respectively. (b) Cross section of the waveguide. (c) Enlarged partial view of a small segment on the boundary of the waveguide core assumed to undergo a small displacement $\boldsymbol{u}$.

As shown in Fig. 1(a), we consider a waveguide that is translationally invariant in the $z$ direction. Assume that the pump and Stokes waves involved in a SBS process in the considered waveguide are two optical eigenmodes, whose electric field $\mathbf{E}$ and electric displacement field $\mathbf{D}$ can be written as:

$$\mathbf{E}^{(i)} = \tfrac{1}{2}\mathbf{e}^{(i)} + \text{c.c.} = \tfrac{1}{2}\tilde{\mathbf{e}}^{(i)} e^{j(\beta^{(i)}z - \omega^{(i)}t)} + \text{c.c.}, \tag{1}$$

$$\mathbf{D}^{(i)} = \tfrac{1}{2}\mathbf{d}^{(i)} + \text{c.c.} = \tfrac{1}{2}\tilde{\mathbf{d}}^{(i)} e^{j(\beta^{(i)}z - \omega^{(i)}t)} + \text{c.c.}. \tag{2}$$

Here, the subscripts $i=1$ or 2 with 1 denoting the Stokes wave and 2 the pump wave; $j = \sqrt{-1}$; $\beta$ and $\omega$ are the wave number and angular frequency of the optical wave, respectively; $t$ is time; and "c.c." is the abbreviation for "complex conjugation". In SBS, the optical fields are perturbed due to the PE and MI effects, and consequently their complex amplitudes $\tilde{\mathbf{e}}^{(i)}$ and $\tilde{\mathbf{d}}^{(i)}$ will generally be modulated by envelope functions varying slowly with $z$ and $t$.

On the acoustic part of the problem, we assume that an elastic wave in solids is involved. According to the well-known non-propagating phonon approximation [1~3], the elastic fields at any position $z$ can essentially be viewed as the local mechanical responses of the waveguide system to the optical forces induced by the electrostriction and radiation pressure effects (Note that in the present study we do not consider the situations in which the enslavement of acoustic wave may break down [34, 39, 40]). Assume that the total optical force density can be written as:

$$\mathbf{F} = \tfrac{1}{2}\mathbf{f} + \text{c. c.} = \tfrac{1}{2}\tilde{\mathbf{f}}e^{j(qz-\Omega t)} + \text{c. c.} \tag{3}$$

Here, the angular frequency $\Omega$ and wavenumber $q$ should satisfy the following phase matching conditions

$$\Omega = \omega^{(2)} - \omega^{(1)}, \tag{4a}$$
$$q = \beta^{(2)} - \beta^{(1)}. \tag{4b}$$

Then, the displacement field induced by the optical forces could be defined in the following form

$$\boldsymbol{\mathcal{U}} = \tfrac{1}{2}\boldsymbol{u} + \text{c. c.} = \tfrac{1}{2}\tilde{\boldsymbol{u}}e^{j(qz-\Omega t)} + \text{c. c.}. \tag{5}$$

Here, we note that a different font style is used for denoting the optically induced displacement field to differentiate it from the intrinsic elastic eigenmode of the waveguide mentioned later. Given the incident optical fields, the finite-element based computational methods for the resultant elastic fields are detailed in Refs. [9, 10, 18].

## 2.2 SBS Gain and acousto-optical coupling coefficients

The growth of the Stokes wave in a steady-state SBS process can generally be described by the following canonical equation [16]:

$$\frac{dP^{(1)}}{dz} = G_B P^{(2)} P^{(1)} - \alpha^{(1)} P^{(1)}. \tag{6}$$

Here, $P^{(1)}$ and $P^{(2)}$ are the $z$-dependent guided powers of the Stokes and pump waves, respectively; $G_B$ is the SBS gain; and $\alpha^{(1)}$ is the linear loss-factor of the Stokes wave. According to Refs. [16~18], the SBS gain can be expressed as:

$$G_B = \frac{\omega^{(1)}}{2\Omega P^{(2)} P^{(1)}} \operatorname{Re}\left[\int \left(\tilde{\mathbf{f}}^* \cdot \frac{d\tilde{\boldsymbol{u}}}{dt}\right) d^2\mathbf{r}\right] = \frac{\omega^{(1)}}{2P^{(2)} P^{(1)}} \operatorname{Im}(\int \tilde{\mathbf{f}}^* \cdot \tilde{\boldsymbol{u}} d^2\mathbf{r}), \tag{7}$$

Here, the integrals are defined over the whole transverse plane of the waveguide.

If the elastic wavenumber $q$ is determined, a set of elastic eigenmodes of the waveguide with different frequencies can be solved. By decomposing the optically induce displacement field in terms of such set of elastic eigenmodes, the SBS gain $G_B$ can be further approximated by a weighted sum of the SBS gains of individual elastic eigenmodes as [18]

$$G_B = \sum_m G_m \frac{(\Gamma_m/2)^2}{(\Omega-\Omega_m)^2 + (\Gamma_m/2)^2}. \tag{8}$$

Here, $\Omega_m$ is the angular frequency of the $m^{\text{th}}$ elastic eigenmode of the corresponding mechanically lossless waveguide, which is changed to a complex value $\Omega_m - j\Gamma_m/2$ if the first-order effect of mechanical loss is considered; and the SBS gain $G_m$ of the $m^{\text{th}}$ elastic eigenmode can be expressed as

$$G_m = \frac{\omega^{(1)} Q_m |C_m|^2}{4P^{(2)} P^{(1)} \mathcal{E}_m}. \tag{9}$$

In Eq. (9), $C_m = \int (\tilde{\mathbf{f}}^* \cdot \tilde{\mathbf{u}}_m) d^2\mathbf{r}$, $Q_m = \Omega_m/\Gamma_m$ and $\mathcal{E}_m = \Omega_m^2/2 \int \rho |\tilde{\mathbf{u}}_m|^2 d^2\mathbf{r}$ are the AOCC, the mechanical $Q$-factor and the average elastic energy per unit length of

the $m^{\text{th}}$ elastic eigenmode, respectively.

The AOCC given above is defined in terms of optical forces. It can be further decomposed into the contributing parts corresponding to the electrostriction and radiation pressure effects. Alternatively, the AOCC can also be defined in terms of the converse physical effects, i.e. the PE and MI effects. According to the analysis in Ref. [36], if the irreversible coupling effects are negligible and the optical losses are sufficiently small, we should have

$$C = C_{\text{ES}} + C_{\text{RP}} = C_{\text{PE}} + C_{\text{MI}}, \tag{10}$$

Here, $C_{\text{ES}}$ and $C_{\text{RP}}$ are the AOCCs due to electrostriction and radiation pressure effects, respectively, and $C_{\text{PE}}$ and $C_{\text{MI}}$ are the ones due to the PE and MI effects, respectively. In the present study, we use Eqs. (8) and (9) to calculate the SBS gain. However, based on the relation (10), the total AOCC of each elastic eigenmode present in Eq. (9), will be calculated as the sum of $C_{\text{PE}}$ and $C_{\text{MI}}$ rather than $C_{\text{ES}}$ and $C_{\text{RP}}$.

If the waveguide materials are optically isotropic, then the AOCCs $C_{\text{PE}}$ and $C_{\text{MI}}$ can be expressed, respectively, as the following surface and contour integrals [26, 33, 36, 37]:

$$C_{\text{PE}} = -\varepsilon_0 \int \varepsilon_r^2 \left(\tilde{\mathbf{e}}^{(1)}\right)^* \cdot \left[(\mathbf{p} : \tilde{\mathbf{s}}^*) \cdot \tilde{\mathbf{e}}^{(2)}\right] d^2 \mathbf{r}, \tag{11}$$

$$C_{\text{MI}} = \oint_\Sigma \left[\left(\varepsilon^{\text{I}} - \varepsilon^{\text{II}}\right)\left(\tilde{\mathbf{e}}_\parallel^{(1)}\right)^* \cdot \tilde{\mathbf{e}}_\parallel^{(2)} - \left(\frac{1}{\varepsilon^{\text{I}}} - \frac{1}{\varepsilon^{\text{II}}}\right)\left(\tilde{\mathbf{d}}_\perp^{(1)}\right)^* \cdot \tilde{\mathbf{d}}_\perp^{(2)}\right] [\tilde{\mathbf{u}}^* \cdot \hat{\mathbf{n}}_\perp] \, dl. \tag{12}$$

The contour integral is defined on the boundary of the waveguide core (denoted by "Σ" in Fig. 1 (b)). In Eq. (11), $\varepsilon_0$ is the permittivity of vacuum, $\varepsilon_r$ is the relative permittivity, **p** is the PE tensor, and **s** is the strain tensor. In Eq. (12), $\varepsilon^{\text{I}}$ and $\varepsilon^{\text{II}}$ are the permitivities of the materials I and II (see Fig. 1), respectively; the subscripts "⊥" and "∥" refer to the normal and tangential components of a field vector on the boundary; and $\hat{\mathbf{n}}_\perp$ is the unit normal vector. As a note, the two contour formulae given in different references may differ by a scaling factor due to different normalization criteria.

### 3. PE and MI acousto-optical coupling coefficients in an anisotropic waveguide

#### 3.1 General integral representations of acousto-optical coupling coefficients

In this section, Eqs. (11) and (12) will be extended to an optically anisotropic waveguide. To this end, the following general integral representations of the AOCCs [36],

$$C_{\text{PE}} = \int \left[\left(\tilde{\mathbf{e}}^{(1)}\right)^* \cdot \left(\Delta \tilde{\mathbf{d}}_{\text{PE}}^{(1)}\right)\right] d^2 \mathbf{r}, \tag{13}$$

$$C_{\text{MI}} = \oint_\Sigma \left[\Delta \tilde{\mathbf{d}}_{\parallel,\text{MI}}^{(2)} \cdot \left(\tilde{\mathbf{e}}_\parallel^{(1)}\right)^* - \Delta \tilde{\mathbf{e}}_{\perp,\text{MI}}^{(2)} \cdot \left(\tilde{\mathbf{d}}_\perp^{(1)}\right)^*\right] [(\tilde{\mathbf{u}})^* \cdot \hat{\mathbf{n}}_\perp] \, dl, \tag{14}$$

will serve as the basis of our derivations. In Eq. (13), $\Delta \mathbf{d}_{\text{PE}}^{(1)}$ is the PE perturbation on the electric displacement field of the Stokes wave. In Eq. (14), $\Delta \mathbf{d}_{\parallel,\mathrm{MI}}$ and $\Delta \mathbf{e}_{\perp,\mathrm{MI}}$ are the changes of the discontinuous optical field components $\mathbf{d}_{\parallel}$ and $\mathbf{e}_{\perp}$ across the boundary, respectively. As shown in Fig. 1(c), they are caused by a small displacement of the boundary. It should be noted that Eq. (14) is not directly given in Ref. [36]. However, it is not difficult to derive it by following the derivation procedures presented in Sections. III.A and III.D of Ref. [36] and simultaneously putting no restrictions on whether the materials are isotropic or not.

**3.2 PE acousto-optical coupling coefficient**

Let's start from the following basic constitutive relation

$$\mathbf{D} = \varepsilon_0 \boldsymbol{\varepsilon}_r \cdot \mathbf{E} \quad (15\mathrm{a})$$
$$= \varepsilon_0 \boldsymbol{\eta}^{-1} \cdot \mathbf{E}, \quad (15\mathrm{b})$$

where $\boldsymbol{\varepsilon}_r$ is the relative permittivity tensor and $\boldsymbol{\eta}$ the impermeability tensor. At the presence of the strain field $\mathbf{s}$, the PE effect results in a small change of the impermeability tensor expressed as [52]:

$$\Delta \boldsymbol{\eta} = \mathbf{p} : \mathbf{s}. \quad (16)$$

Then, we have

$$\Delta(\boldsymbol{\eta}^{-1}) = (\boldsymbol{\varepsilon}_r^{-1} + \mathbf{p}:\mathbf{s})^{-1} - (\boldsymbol{\varepsilon}_r^{-1})^{-1} \quad (17\mathrm{a})$$
$$= [\boldsymbol{\varepsilon}_r^{-1}(\mathbf{I} + \boldsymbol{\varepsilon}_r \cdot (\mathbf{p}:\mathbf{s}))]^{-1} - \boldsymbol{\varepsilon}_r \quad (17\mathrm{b})$$
$$\approx -\boldsymbol{\varepsilon}_r \cdot (\mathbf{p}:\mathbf{s}) \cdot \boldsymbol{\varepsilon}_r, \quad (17\mathrm{c})$$

where $\mathbf{I}$ is the unity tensor.

Consequently, the change of the electric displacement field $\mathbf{D}$ caused by the PE effect can be written as

$$\Delta \mathbf{D} = -\varepsilon_0 [\boldsymbol{\varepsilon}_r \cdot (\mathbf{p}:\mathbf{s}) \cdot \boldsymbol{\varepsilon}_r] \cdot \mathbf{E}. \quad (18)$$

Then, the change of the electric displacement field phase-matched with the Stokes wave, i.e. $\Delta \mathbf{d}_{\mathrm{PE}}^{(1)}$, can be written as:

$$\Delta \mathbf{d}_{\mathrm{PE}}^{(1)} = -\varepsilon_0 [\boldsymbol{\varepsilon}_r \cdot (\mathbf{p}:\mathbf{s}^*) \cdot \boldsymbol{\varepsilon}_r] \cdot \mathbf{e}^{(2)}. \quad (19)$$

Rewriting Eq. (19) by replacing the fields $\mathbf{d}$, $\mathbf{s}$ and $\mathbf{e}$ with their complex amplitudes, and then substituting it into Eq. (13), we could eventually derive the AOCC due to the PE effect in an anisotropic waveguide:

$$C_{\mathrm{PE}} = -\varepsilon_0 \int (\tilde{\mathbf{e}}^{(1)})^* \cdot \{[\boldsymbol{\varepsilon}_r \cdot (\mathbf{p}:\tilde{\mathbf{s}}^*) \cdot \boldsymbol{\varepsilon}_r] \cdot \tilde{\mathbf{e}}^{(2)}\} d^2 \mathbf{r} \quad (20)$$

The correctness of Eq. (20) can be checked by assuming that the tensor $\boldsymbol{\varepsilon}_r = \varepsilon_r \mathbf{I}$. Then, Eq. (20) expectedly becomes Eq. (11) in the isotropic case.

### 3.3 MI acousto-optical coupling coefficient

The extension of the MI AOCC to an anisotropic waveguide is not as straightforward as that for the PE AOCC presented above. The first and crucial step to do it is to represent the discontinuous field components $\mathbf{E}_\perp$ and $\mathbf{D}_\parallel$ in terms of the continuous field components $\mathbf{E}_\parallel$ and $\mathbf{D}_\perp$. To this end, we establish an in-plane Cartesian coordinate system $x'$-$y'$ at any point at the interface. As shown in Fig. 1(b), the axes $x'$ and $y'$ are along the locally tangential and normal directions, respectively.

In the local coordinate system, the electric field $\mathbf{E}$ and electric displacement field $\mathbf{D}$ can be decomposed as

$$\mathbf{E} = \mathbf{E}_{\parallel,x'} + \mathbf{E}_{\parallel,z} + \mathbf{E}_{y'}, \tag{21}$$
$$\mathbf{D} = \mathbf{D}_{\parallel,x'} + \mathbf{D}_{\parallel,z} + \mathbf{D}_{y'}, \tag{22}$$

Then, according to the basic constitutive relation $\mathbf{D} = \boldsymbol{\varepsilon} \cdot \mathbf{E}$, we have

$$D_{\parallel,x'} - \varepsilon_{x'y'} E_{y'} = \varepsilon_{x'x'} E_{\parallel,x'} + \varepsilon_{x'z} E_{\parallel,z}, \tag{23a}$$
$$D_{\parallel,z} - \varepsilon_{zy'} E_{y'} = \varepsilon_{zx'} E_{\parallel,x'} + \varepsilon_{zz} E_{\parallel,z}, \tag{23b}$$
$$\varepsilon_{y'y'} E_{y'} = -\varepsilon_{y'x'} E_{\parallel,x'} - \varepsilon_{y'z} E_{\parallel,z} + D_{y'}. \tag{23c}$$

Rewriting Eq. (23) in a compact matrix form, the discontinuous field components $D_{\parallel,x'}$, $D_{\parallel,z}$ and $E_{y'}$ can be represented in terms of the continuous ones $E_{\parallel,x'}$, $E_{\parallel,z}$ and $D_{y'}$ as:

$$\begin{pmatrix} D_{\parallel,x'} \\ D_{\parallel,z} \\ E_{y'} \end{pmatrix} = \boldsymbol{\zeta} \begin{pmatrix} E_{\parallel,x'} \\ E_{\parallel,z} \\ D_{y'} \end{pmatrix}, \tag{24}$$

where

$$\boldsymbol{\zeta} = \begin{bmatrix} 1 & 0 & -\varepsilon_{x'y'} \\ 0 & 1 & -\varepsilon_{zy'} \\ 0 & 0 & \varepsilon_{y'y'} \end{bmatrix}^{-1} \begin{bmatrix} \varepsilon_{x'x'} & \varepsilon_{x'z} & 0 \\ \varepsilon_{zx'} & \varepsilon_{zz} & 0 \\ -\varepsilon_{y'x'} & -\varepsilon_{y'z} & 1 \end{bmatrix} \tag{25a}$$

$$= \begin{bmatrix} \varepsilon_{x'x'} - (\varepsilon_{x'y'})^2/\varepsilon_{y'y'} & \varepsilon_{x'z} - \varepsilon_{x'y'}\varepsilon_{zy'}/\varepsilon_{y'y'} & \varepsilon_{x'y'}/\varepsilon_{y'y'} \\ \varepsilon_{x'z} - \varepsilon_{x'y'}\varepsilon_{zy'}/\varepsilon_{y'y'} & \varepsilon_{zz} - (\varepsilon_{zy'})^2/\varepsilon_{y'y'} & \varepsilon_{zy'}/\varepsilon_{y'y'} \\ -\varepsilon_{x'y'}/\varepsilon_{y'y'} & -\varepsilon_{zy'}/\varepsilon_{y'y'} & 1/\varepsilon_{y'y'} \end{bmatrix}. \tag{25b}$$

In the step from Eqs. (25a) to (25b), the symmetry of the permittivity tensor $\boldsymbol{\varepsilon}$ ($\varepsilon_{ij} = \varepsilon_{ji}$) is used.

Here, we would like to emphasize that the $3 \times 3$ matrix $\boldsymbol{\zeta}$ has the following interesting properties:

$$\zeta_{12} = \zeta_{21}, \tag{26a}$$
$$\zeta_{13} = -\zeta_{31}, \tag{26b}$$
$$\zeta_{23} = -\zeta_{32}, \tag{26c}$$

which will be essential in obtaining the final derivation result.

To highlight the final derivation result, the detailed derivation subsequent to Eq. (26) is put in Appendix A. The final contour-integral formula for the MI OACC can be written as:

$$C_{\mathrm{MI}} =$$

$$\oint_\Sigma \left\{\left[\varepsilon^{\mathrm{I}}_{x'x'} - \varepsilon^{\mathrm{II}}_{x'x'} - \left(\varepsilon^{\mathrm{I}}_{x'y'}\right)^2/\varepsilon^{\mathrm{I}}_{y'y'} + \left(\varepsilon^{\mathrm{II}}_{x'y'}\right)^2/\varepsilon^{\mathrm{II}}_{y'y'}\right]\left[\tilde{e}^{(2)}_{\parallel,x'}\left(\tilde{e}^{(1)}_{\parallel,x'}\right)^*\right] + \left[\varepsilon^{\mathrm{I}}_{zz} - \varepsilon^{\mathrm{II}}_{zz} - \right.\right.$$

$$\left(\varepsilon^{\mathrm{I}}_{y'z}\right)^2/\varepsilon^{\mathrm{I}}_{y'y'} + \left(\varepsilon^{\mathrm{II}}_{y'z}\right)^2/\varepsilon^{\mathrm{II}}_{y'y'}\right]\left[\tilde{e}^{(2)}_{\parallel,z}\left(\tilde{e}^{(1)}_{\parallel,z}\right)^*\right] - \left(1/\varepsilon^{\mathrm{I}}_{y'y'} - 1/\varepsilon^{\mathrm{II}}_{y'y'}\right)\left[\tilde{d}^{(2)}_{y'}\left(\tilde{d}^{(1)}_{y'}\right)^*\right] +$$

$$\left(\varepsilon^{\mathrm{I}}_{x'y'}/\varepsilon^{\mathrm{I}}_{y'y'} - \varepsilon^{\mathrm{II}}_{x'y'}/\varepsilon^{\mathrm{II}}_{y'y'}\right)\left[\tilde{e}^{(2)}_{\parallel,x'}\left(\tilde{d}^{(1)}_{y'}\right)^* + \left(\tilde{e}^{(1)}_{\parallel,x'}\right)^*\tilde{d}^{(2)}_{y'}\right] +$$

$$\left(\varepsilon^{\mathrm{I}}_{y'z}/\varepsilon^{\mathrm{I}}_{y'y'} - \varepsilon^{\mathrm{II}}_{y'z}/\varepsilon^{\mathrm{II}}_{y'y'}\right)\left[\tilde{e}^{(2)}_{\parallel,z'}\left(\tilde{d}^{(1)}_{y'}\right)^* + \left(\tilde{e}^{(1)}_{\parallel,z'}\right)^*\tilde{d}^{(2)}_{y'}\right] + \left(\varepsilon^{\mathrm{I}}_{x'z} - \varepsilon^{\mathrm{II}}_{x'z} - \varepsilon^{\mathrm{I}}_{x'y'}\varepsilon^{\mathrm{I}}_{y'z}/\varepsilon^{\mathrm{I}}_{y'y'} + \right.$$

$$\left.\left.\varepsilon^{\mathrm{II}}_{x'y'}\varepsilon^{\mathrm{II}}_{y'z}/\varepsilon^{\mathrm{II}}_{y'y'}\right)\left[\tilde{e}^{(2)}_{\parallel,z}\left(\tilde{e}^{(1)}_{\parallel,x'}\right)^* + \tilde{e}^{(2)}_{\parallel,x'}\left(\tilde{e}^{(1)}_{\parallel,z}\right)^*\right]\right\}\left[(\tilde{\mathbf{u}})^* \cdot \hat{\mathbf{n}}_{y'}\right] dl. \tag{27}$$

The obtained contour-integral is much more complex than in the isotropic case. However, it can be checked that if the dielectric tensors $\boldsymbol{\varepsilon}^{\mathrm{I}}$ and $\boldsymbol{\varepsilon}^{\mathrm{II}}$ both reduce to scalars, Eq. (27) will reduce to Eq. (12) in the isotropic case accordingly.

### 3.4 Simplified formulae in an orthotropic case

So far, we have derived the general integral formulae for the PE and MI AOCCs in an optically anisotropic waveguide. Next, let's consider a special case that the waveguide materials are both orthotropic and simultaneously they share a common set of principal axes. For convenience, the geometrical axes (the $x$-, $y$- and $z$-axes shown in Fig. 1(a)) are assumed to coincide with the common principal axes of the two materials. For such a special case, we will have $\varepsilon_{ij} = \varepsilon_i \delta_{ij}$ in the global coordinate system and $\varepsilon_{x'z} = \varepsilon_{y'z} = 0$ in the in-plane local coordinate system defined in Fig. 1(b). Then, Eqs. (20) and (27) can be further simplified as:

$$C_{\mathrm{PE}} = -\varepsilon_0 \int \left[\varepsilon_{\mathrm{r},i}\varepsilon_{\mathrm{r},j}\left(\tilde{e}^{(1)}_i\right)^* \tilde{e}^{(2)}_j p_{ijlm} \tilde{s}^*_{lm}\right] d^2\mathbf{r} \tag{28}$$

$$C_{\mathrm{MI}} =$$

$$\oint_\Sigma \left\{\left[\varepsilon^{\mathrm{I}}_{x'x'} - \varepsilon^{\mathrm{II}}_{x'x'} - \left(\varepsilon^{\mathrm{I}}_{x'y'}\right)^2/\varepsilon^{\mathrm{I}}_{y'y'} + \left(\varepsilon^{\mathrm{II}}_{x'y'}\right)^2/\varepsilon^{\mathrm{II}}_{y'y'}\right]\left[\tilde{e}^{(2)}_{\parallel,x'}\left(\tilde{e}^{(1)}_{\parallel,x'}\right)^*\right] +\right.$$

$$\left[\varepsilon^{\mathrm{I}}_z - \varepsilon^{\mathrm{II}}_z\right]\left[\tilde{e}^{(2)}_{\parallel,z}\left(\tilde{e}^{(1)}_{\parallel,z}\right)^*\right] - \left(1/\varepsilon^{\mathrm{I}}_{y'y'} - 1/\varepsilon^{\mathrm{II}}_{y'y'}\right)\left[\tilde{d}^{(2)}_{y'}\left(\tilde{d}^{(1)}_{y'}\right)^*\right] + \left(\varepsilon^{\mathrm{I}}_{x'y'}/\varepsilon^{\mathrm{I}}_{y'y'} - \right.$$

$$\left.\left.\varepsilon^{\mathrm{II}}_{x'y'}/\varepsilon^{\mathrm{II}}_{y'y'}\right)\left[\tilde{e}^{(2)}_{\parallel,x'}\left(\tilde{d}^{(1)}_{y'}\right)^* + \left(\tilde{e}^{(1)}_{\parallel,x'}\right)^*\tilde{d}^{(2)}_{y'}\right]\right\}\left[(\tilde{\mathbf{u}})^* \cdot \hat{\mathbf{n}}_{y'}\right] dl. \tag{29}$$

For the calculation of $C_{\text{MI}}$ via Eq. (29), we need to further substitute the following expressions:

$$\varepsilon_{x'x'} = n_y^2 \varepsilon_x + n_x^2 \varepsilon_y, \tag{30a}$$

$$\varepsilon_{y'y'} = n_x^2 \varepsilon_x + n_y^2 \varepsilon_y, \tag{30b}$$

$$\varepsilon_{x'y'} = (\varepsilon_x - \varepsilon_y) n_x n_y, \tag{30c}$$

$$\tilde{e}_{\parallel,x'} = \tilde{e}_x n_y - \tilde{e}_y n_x, \tag{30d}$$

$$\tilde{d}_{y'} = \tilde{d}_x n_x + \tilde{d}_y n_y, \tag{30e}$$

$$(\tilde{\mathbf{u}})^* \cdot \hat{\mathbf{n}}_{y'} = \tilde{u}_x n_x + \tilde{u}_y n_y, \tag{30f}$$

Here, $n_x$ and $n_y$ are the directional cosines of the normal vector $\hat{\mathbf{n}}_{y'}$. They can be expressed as $n_x = -\sin(\theta)$ and $n_y = \cos(\theta)$ with the angle $\theta$ being defined in Figure. 1(b).

In the finite-element software COMSOL Multiphysics, the calculations of $C_{\text{PE}}$ and $C_{\text{MI}}$ can be implemented via the built-in integration operators. Moreover, with the built-in variables "nx" and "ny" for defining a surface normal, the comparably complex contour integral calculation of $C_{\text{MI}}$ can be implemented much more conveniently.

In the computational aspect, there is another important issue needed to be addressed. In a longitudinally invariant waveguide, there is generally a fixed phase difference of π/2 between the longitudinal field component and the transverse ones for both kinds of waves. By adjusting the initial phases of the transverse field components to zero, the transverse and longitudinal field components will become purely real-valued and purely imaginary-valued, respectively. Consequently, just like the isotropic case [26, 35], it can be checked that the PE and MI AOCCs in the orthotropic case given by Eqs. (28) and (29) can both reduce to pure real numbers.

## 4. Numerical simulations

### 4.1 Simulation settings

As shown in Fig. 2, an elliptical nanowire is considered in the numerical simulations. As discussed in Sec. 3.4, the waveguide material is assumed to be orthotropic. The geometrical axes *x*, *y* and *z* are aligned with the principal material axes. But in the transverse plane, the symmetry axes of the ellipse (i.e. waveguide axes) are not assumed to be aligned with the principal material axes. There is an angle γ between the major axis of the ellipse and the *x*-axis. Therefore, if the material property is transversely anisotropic, the orientation angle γ should have effect on propagation properties of both the two kinds of waves and their interactions as well. In the present study, we focus on the effects of the orientation angle γ on the SBS gains and AOCCs.

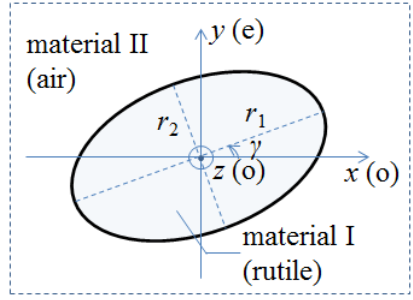

Fig. 2 Transverse geometry of the considered elliptical nanowire

The orthotropic material we select is the single-crystalline rutile (titanium dioxide), which exhibits a strong birefringence. With many notable merits, such as the high refractive index, large optical nonlinearity and transparency at visible and near-infrared wavelength, rutile has become a promising material for integrated optics in recent years [53~55]. The operating vacuum optical wavelength is selected as 633 nm, as the optical parameters of rutile are usually available at this wavelength in references. With the semi-major axis $r_1=140$ nm and semi-minor axis $r_2=90$ nm, the transverse dimensions of the considered waveguide fall well within the sub-wavelength regime.

As shown in Fig. 2, an extraordinary axis of the material is adopted as the *y*-axis, and then the *x*- and *z*-axes are two ordinary axes of the material. Under such definitions of the geometrical axes, the desired transverse anisotropy is enabled. The physical parameters used for rutile are listed in Ref. [56]. These parameters are originally taken from Refs. [52] and [57]. Additionally, a frequency-independent mechanical *Q*-factor of 1000 is set for the elastic wave. Here, we note that compared with the original references, different abbreviated subscripts may be used for some components of material property tensors, as the definitions of geometric axes do not follow general convention. For a uniaxial material, the extraordinary material axis is conventionally adopted as the *z*-axis rather than the *y*-axis in the present study.

In the present study, the intra-mode SBS induced by the fundamental optical mode is considered. The optical wavenumber $\beta$ is searched according to the assumed wavelength 633 nm. The elastic wavenumber $q$ is further determined as $q=0$ and $2\beta$ for FSBS and backward SBS (BSBS), respectively. Then, for the given wavenumber $q$, we compute the elastic eigenmodes, via which the PE and MI AOCCs at different phononic eigenfrequencies are further computed. Finally, the SBS gain at any given phononic frequency is calculated via Eq. (8). The finite-element software COMSOL Multiphysics is used for the computations.

For convenience, the initial guided powers of optical waves as well as the energy of each elastic eigenmode will all be normalized to unity in the computations. In the normalization, the optical guided power is calculated as $P = \frac{1}{2}\int \text{Re}\{\hat{z} \cdot [(\tilde{\mathbf{e}})^* \times$

$\tilde{\mathbf{h}}]\} d^2\mathbf{r}$ with $\mathbf{h}$ the magnetic field. After the normalization, we should have $\tilde{\mathbf{e}}^{(1)} = \tilde{\mathbf{e}}^{(2)}$ and $\tilde{\mathbf{e}}^{(1)} = (\tilde{\mathbf{e}}^{(2)})^*$ for FSBS and BSBS, respectively. And then, based on the above-mentioned material settings for the waveguide, the expressions of the PE and MI AOCCs can be further simplified. For FSBS, the two coupling coefficients are written as:

$$C_{\text{PE}} = -\varepsilon_0 \int \left[ \varepsilon_{r,x}^2 |\tilde{e}_x|^2 (p_{11}\tilde{s}_{xx}^* + p_{12}\tilde{s}_{yy}^*) + \varepsilon_{r,y}^2 |\tilde{e}_y|^2 (p_{21}\tilde{s}_{xx}^* + p_{22}\tilde{s}_{yy}^*) + \varepsilon_{r,z}^2 |\tilde{e}_z|^2 (p_{31}\tilde{s}_{xx}^* + p_{32}\tilde{s}_{yy}^*) \right] d^2\mathbf{r}, \tag{31}$$

$$C_{\text{MI}} =$$
$$\oint_\Sigma \left\{ \varepsilon_0 \left[ \left( n_y^2 \varepsilon_{r,x} + n_x^2 \varepsilon_{r,y} - 1 - \left( (\varepsilon_{r,x} - \varepsilon_{r,y}) n_x n_y \right)^2 / (n_x^2 \varepsilon_{r,x} + n_y^2 \varepsilon_{r,y}) \right) |\tilde{e}_x n_y - \tilde{e}_y n_x|^2 + (\varepsilon_{r,z} - 1)|\tilde{e}_z|^2 \right] - \frac{1}{\varepsilon_0} \left[ (1/(n_x^2 \varepsilon_{r,x} + n_y^2 \varepsilon_{r,y}) - 1) |\tilde{d}_x n_x + \tilde{d}_y n_y|^2 \right] + 2 \left( (\varepsilon_{r,x} - \varepsilon_{r,y}) n_x n_y \right) / (n_x^2 \varepsilon_{r,x} + n_y^2 \varepsilon_{r,y}) \cdot \text{Re} \left( (\tilde{e}_x n_y - \tilde{e}_y n_x)(\tilde{d}_x n_x + \tilde{d}_y n_y)^* \right) \right\} \left[ (\tilde{u}_x)^* n_x + (\tilde{u}_y)^* n_y \right] dl \tag{32}$$

And for BSBS, they are written as:

$$C_{\text{PE}} = -\varepsilon_0 \int \left[ \varepsilon_{r,x}^2 (\tilde{e}_x)^2 (p_{11}\tilde{s}_{xx}^* + p_{12}\tilde{s}_{yy}^* + p_{13}\tilde{s}_{zz}^*) + \varepsilon_{r,y}^2 (\tilde{e}_y)^2 (p_{21}\tilde{s}_{xx}^* + p_{22}\tilde{s}_{yy}^* + p_{23}\tilde{s}_{zz}^*) + \varepsilon_{r,z}^2 (\tilde{e}_z)^2 (p_{31}\tilde{s}_{xx}^* + p_{32}\tilde{s}_{yy}^* + p_{33}\tilde{s}_{zz}^*) + 4\varepsilon_{r,x}\varepsilon_{r,z} \tilde{e}_x \tilde{e}_z p_{55} \tilde{s}_{xz}^* \right] d^2\mathbf{r} \tag{33}$$

$$C_{\text{MI}} =$$
$$\oint_\Sigma \left\{ \varepsilon_0 \left[ \left( n_y^2 \varepsilon_{r,x} + n_x^2 \varepsilon_{r,y} - 1 - \left( (\varepsilon_{r,x} - \varepsilon_{r,y}) n_x n_y \right)^2 / (n_x^2 \varepsilon_{r,x} + n_y^2 \varepsilon_{r,y}) \right) (\tilde{e}_x n_y - \tilde{e}_y n_x)^2 + (\varepsilon_{r,z} - 1)(\tilde{e}_z)^2 \right] - \frac{1}{\varepsilon_0} \left[ (1/(n_x^2 \varepsilon_{r,x} + n_y^2 \varepsilon_{r,y}) - 1)(\tilde{d}_x n_x + \tilde{d}_y n_y)^2 \right] + 2 \left( (\varepsilon_{r,x} - \varepsilon_{r,y}) n_x n_y \right) / (n_x^2 \varepsilon_{r,x} + n_y^2 \varepsilon_{r,y}) (\tilde{e}_x n_y - \tilde{e}_y n_x)(\tilde{d}_x n_x + \tilde{d}_y n_y) \right\} \left[ (\tilde{u}_x)^* n_x + (\tilde{u}_y)^* n_y \right] dl \tag{34}$$

Here, we note that since $p_{44}=p_{66}=0$ and $p_{55} \neq 0$, there are no shear strains appearing in the expression of $C_{MI}$ for FSBS, and only the shear strain $\tilde{s}_{xz}$ appears in the corresponding expression for BSBS.

**4.2 Effects of the waveguide's orientation angle on SBS gain**

Figures 3(a) and (b) show respectively the FSBS and BSBS spectra obtained at three different orientation angles 0°, 45° and 90°. The obtained SBS spectra exhibit typical characteristics in the sub-wavelength regime. There are many resonant peaks corresponding to the elastic eigenmodes of the waveguide. The obtained maximum SBS gain is close to $10^5$ m$^{-1}$W$^{-1}$, which are larger than the previously reported results [16~18] on the order of $10^4$ m$^{-1}$W$^{-1}$ obtained for silicon waveguides with the same mechanical $Q$-factor of 1000. The larger SBS gain can be attributed to a larger optical power density, which is further resulted from the smaller waveguide dimensions suitable for the shorter optical wavelength of 633 nm.

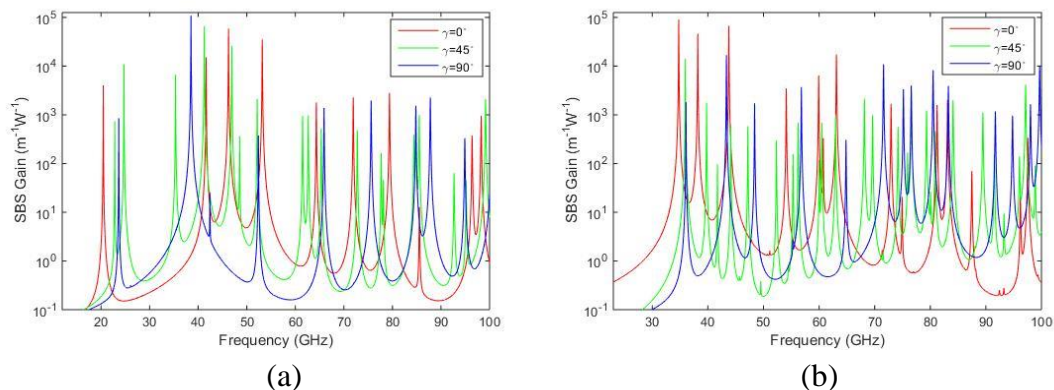

(a)            (b)

Fig. 3 (a) FSBS (b) BSBS gain spectra obtained at three different orientation angles 0°, 45° and 90°.

In Figure 3, it is observed that the resonant phononic frequencies and the corresponding SBS gains are both influenced significantly by the orientation angle γ of the waveguide. The observation is quite understandable. Due to the strong transverse anisotropy of the material property, the elastic eigenfrequencies which determine the resonant phononic frequencies, as well as the elastic and optical eigenmodes which determine the magnitudes of AOCCs, should all be highly dependent on the orientation angle γ.

Next, we report a more interesting phenomenon. Figures. 4(a) and (b) show two neighboring peaks in the gain spectra obtained at five different orientation angles ranging from 0° to 8° for FSBS and BSBS, respectively. It is observed that the first peaks change very little with γ. By contrast, the second peaks are very sensitive to the small change of γ. When γ increases from 0° to just 1°, the gains at the second peaks are increased by approximately 632 and 27 times for FSBS and BSBS, respectively, but the corresponding phononic frequencies change very little; and then, each time the angle γ doubles, the SBS gains nearly quadruple.

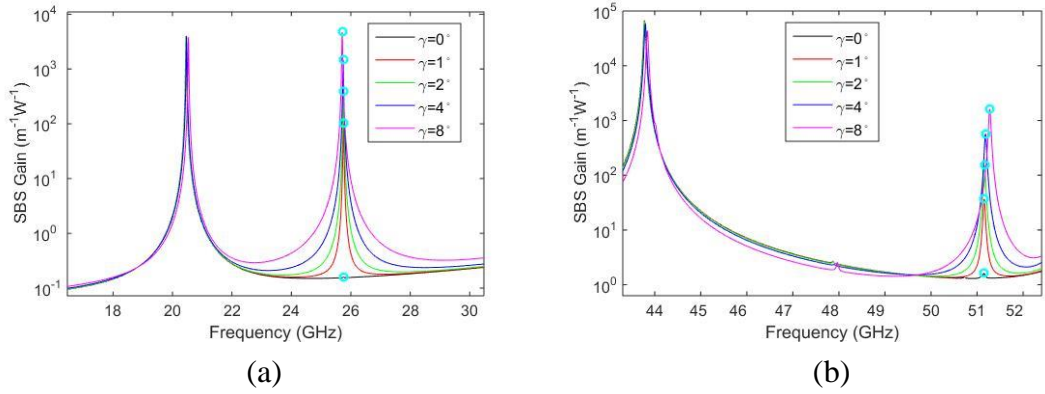

(a)                            (b)

Fig. 4 Two neighboring resonant peaks in the (a) FSBS and (b) BSBS gain spectra obtained at five different orientation angles ranging from 0° to 8°. The second resonant peaks are marked by circles to show the variation with the angle γ.

To understand this interesting phenomenon, we examine the PE and MI AOCCs of the elastic eigenmodes corresponding to the second resonant peaks in Fig. 4. Tables 1 and 2 list these AOCCs as a function of the orientation angle γ in the FSBS and BSBS cases, respectively. It is observed that the AOCCs at γ=0° are at least 100 times smaller than those at γ=1°. We will show later that the AOCCs at γ=0° actually vanish and the obtained non-vanishing results are caused by discretization errors in the computation. Therefore, at γ=0°, the AOCCs could have the largest derivatives with respect to γ. A very small change of γ may result in large changes of AOCCs and SBS gains. Meanwhile, Tables 1 and 2 also show that the AOCCs vary almost linearly with the small angle γ (Note that the linearity weakens gradually with the increase of γ). Consequently, owing to the quadratic dependence of SBS gains on AOCCs (see Eqs. (8) and (9)), doubling the angle γ results in that the SBS gains nearly quadruple. As a note, in spite of the very small AOCCs at γ=0°, the corresponding SBS gains are not so small. In fact, according to Eq. (8), the non-vanishing SBS gains are mainly attributed to the additive contributions from neighboring elastic eigenmodes.

Table 1 Acousto-optical coupling coefficients $C_{PE}$ and $C_{MI}$ as a function of the orientation angle γ in the FSBS case

| γ | 0° | 1° | 2° | 4° | 8° |
|---|---|---|---|---|---|
| $C_{PE}$ (N) | $-3.28\times10^{-12}$ | $1.17\times10^{-9}$ | $2.33\times10^{-9}$ | $4.64\times10^{-9}$ | $9.07\times10^{-9}$ |
| $C_{MI}$ (N) | $6.07\times10^{-11}$ | $7.07\times10^{-9}$ | $1.38\times10^{-8}$ | $2.67\times10^{-8}$ | $4.73\times10^{-8}$ |

Table 2 Acousto-optical coupling coefficients $C_{PE}$ and $C_{MI}$ as a function of the orientation angle γ in the BSBS case

| γ | 0° | 1° | 2° | 4° | 8° |
|---|---|---|---|---|---|
| $C_{PE}$ (N) | $1.73\times10^{-11}$ | $3.87\times10^{-9}$ | $8.09\times10^{-9}$ | $1.58\times10^{-8}$ | $2.61\times10^{-8}$ |
| $C_{MI}$ (N) | $-1.31\times10^{-12}$ | $1.02\times10^{-9}$ | $1.96\times10^{-9}$ | $3.77\times10^{-9}$ | $6.8\times10^{-9}$ |

Besides the ones shown in Figure 4, there are actually many such kind of γ-sensitive resonant peaks in the obtained SBS spectra. Next, we show the SBS spectra in the frequency range below 100 GHz. Figure 5 shows the FSBS and BSBS gain spectra obtained at five different orientation angles ranging from 0° to 8°, and Figure 6 shows the corresponding results for orientation angles ranging from 90° to 98°. The γ-sensitive resonant peaks are all marked by small circles. On average, the percentage of such kind of resonant peaks is as high as 28%. However, if γ is not near 0° or 90°, the γ-sensitive resonant peaks can hardly be found. As Figure 7 shows, at γ=45°, only one resonant peak of such kind is found in the BSBS spectrum. That is to say, the observed interesting phenomenon is much more likely to occur if the waveguide axes are aligned with the principal material axes. The SBS gains of some elastic modes are very sensitive to the small misalignment between the two kinds of axes

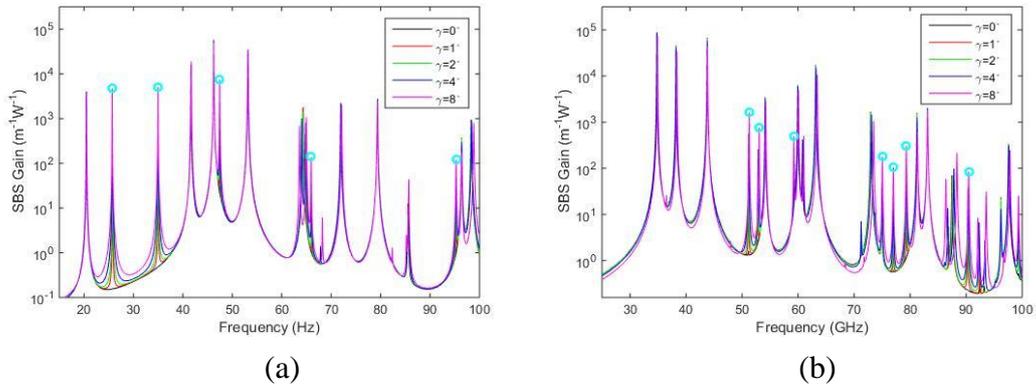

(a)          (b)

Fig. 5 (a) FSBS and (b) BSBS gain spectra obtained at five different orientation angles ranging from 0° to 8°. The γ-sensitive resonant peaks are all marked by small circles.

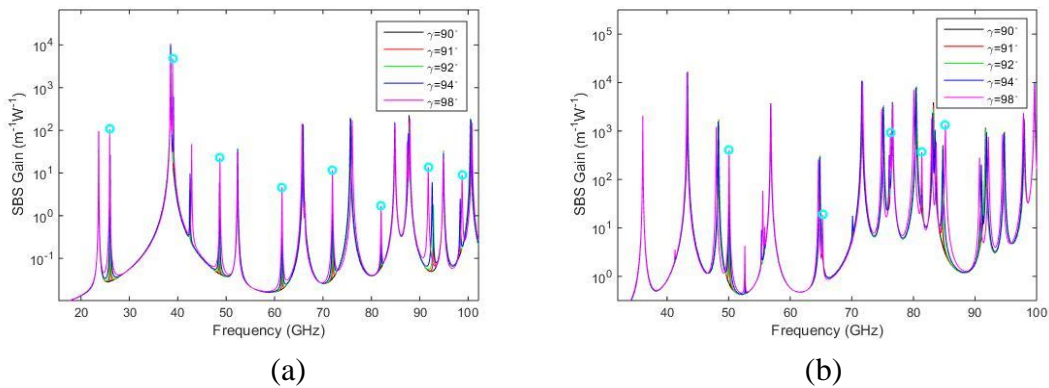

(a)          (b)

Fig. 6 (a) FSBS and (b) BSBS gain spectra obtained at five different orientation angles ranging from 90° to 98°. The γ-sensitive resonant peaks are all marked by small circles.

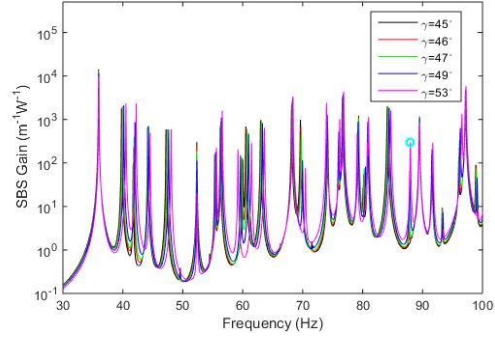

Fig. 7 BSBS gain spectra obtained at five different orientation angles ranging from 45° to 53°. A γ-sensitive resonant peak is marked by small circle.

## 4.3 Detailed modal field analysis

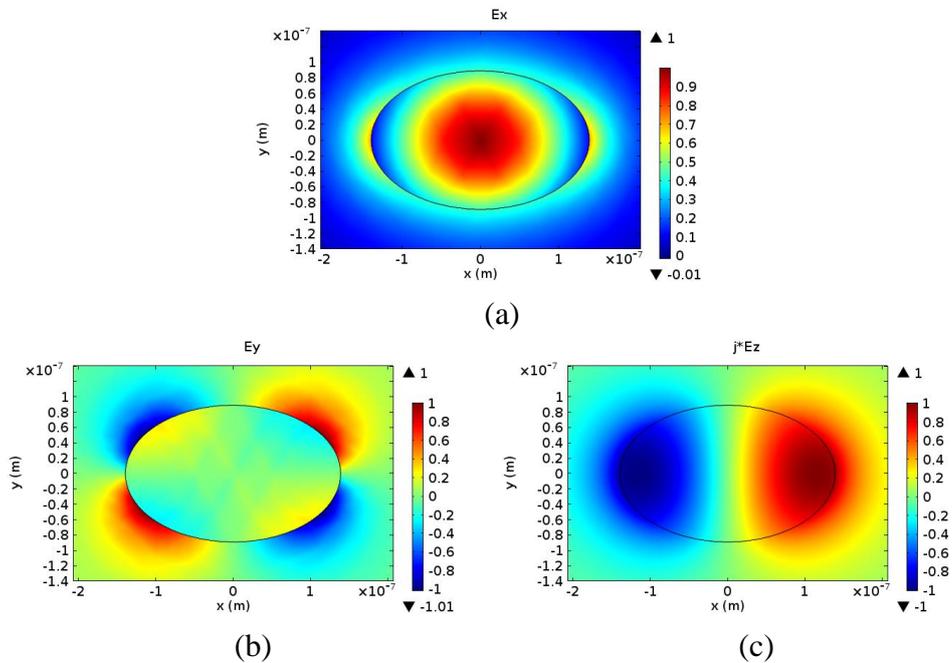

Fig. 8 Spatial profiles of the electric fields (a) $E_x$, (b) $E_y$ and (c) $E_z$ of the optical fundamental mode corresponding to the orientation angle 0°

To gain more insight into the physics behind the obtained results, let's analyze the spatial profiles of the optical and elastic modes involved in SBS. Take the case of γ=0° as an example. Figure 8 shows the spatial profiles of the electric fields of the optical fundamental mode at γ=0°. The fields $E_x$, $E_y$ and $E_z$ all have axisymmetric or anti-axisymmetric structures. $E_x$ ($E_y$) is symmetric (anti-symmetric) about both the x-axis and y-axis, while $E_z$ is symmetric about the x-axis and anti-symmetric about the y-axis. Note that in an orthotropic material, the symmetry property of the electric displacement field coincides completely with that of the electric field. Next, let's first analyze the second resonant peak in the FSBS spectrum shown in Figure 4(a). Figure 9 shows the different field components of the elastic mode at the corresponding eigenfrequency 25.7 GHz. The displacement field $U_x$ ($U_y$) is anti-symmetric (symmetric) about the *x*-axis and symmetric (anti-symmetric) about the *y*-axis, while the strain fields $S_{xx}$ and $S_{yy}$ are both anti-symmetric about the geometrical axes.

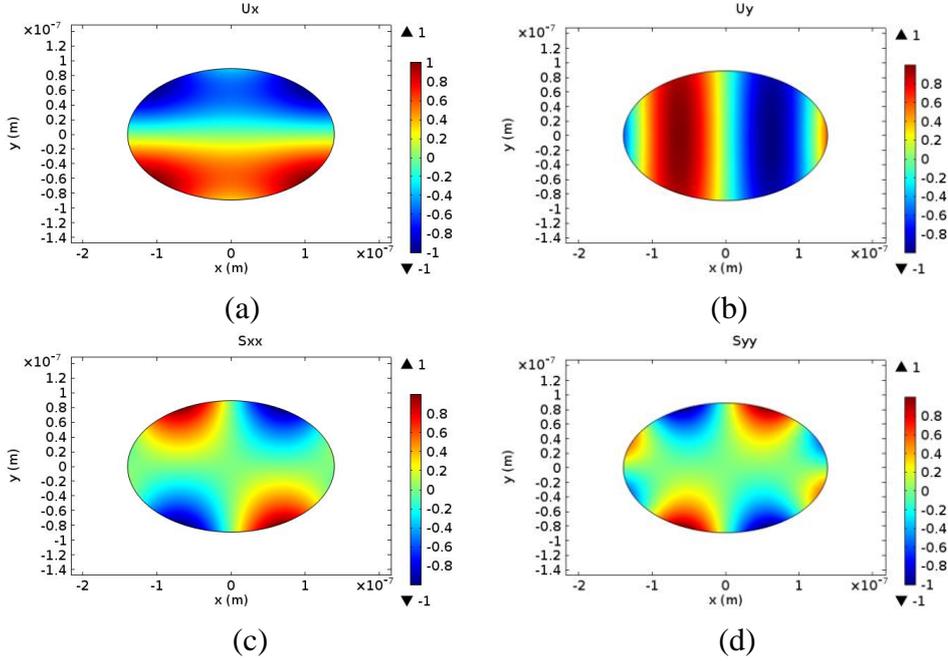

(a)      (b)

(c)      (d)

Fig. 9 Different field components of (a) $U_x$, (b) $U_y$, (c) $S_{xx}$ and (d) $S_{yy}$ of the elastic mode with the eigenfrequency 25.7 GHz corresponding to the case of γ=0° in Fig. 4(a)

Now, let's examine Eq. (31), i.e. the expression of $C_{\text{PE}}$ in the FSBS case. In terms of the axisymmetric structures of $|\tilde{e}_x|$, $|\tilde{e}_y|$ and $|\tilde{e}_z|$, as well as the anti-axisymmetric structure of $\tilde{s}_{xx}$ and $\tilde{s}_{yy}$, it is not difficult to conclude that $C_{\text{PE}}=0$. However, as mentioned above, due to the discretization errors in the computation, the calculated result does not absolutely vanish. The analysis of $C_{\text{MI}}$ is comparably more complex. For the sake of convenience, we rewrite Eq. (32) in the following form:

$$C_{\text{MI}} = \oint_\Sigma A(x,y) \cdot \left[(\tilde{u}_x)^* n_x + (\tilde{u}_y)^* n_y\right] dl. \tag{35}$$

The spatially dependent function *A* is the content inside the brace in Eq. (32), which is further written as an algebraic sum of three parts. Here, the three parts are denoted as $A_1$, $A_2$ and $A_3$, respectively. According to the symmetry properties of the electric fields, as well as those of the directional cosines $n_x$ and $n_y$ ($n_x$ is symmetric about the *x* axis and anti-symmetric about the *y* axis, and $n_y$ reverses), it can be checked that the parts $A_1$, $A_2$ and $A_3$ of *A* are all symmetric about both the two axes. However, the symmetry properties of $\tilde{u}_x$ and $\tilde{u}_y$, as well as those of $n_x$ and $n_y$, jointly result in the anti-symmetry of $\left[(\tilde{u}_x)^* n_x + (\tilde{u}_y)^* n_y\right]$ about the *x* axis. Therefore, the contour integral of $A \cdot \left[(\tilde{u}_x)^* n_x + (\tilde{u}_y)^* n_y\right]$, i.e. $C_{\text{MI}}$, theoretically equals zero.

So far, based on the symmetry properties of the optical and elastic fields, we have obtained that the PE and MI AOCCs corresponding to the second resonant peak in Fig. 4(a) both vanish at γ=0°. To analyze the second resonant peak in the BSBS spectrum shown in Fig. 4(b), we show the different field components of the elastic mode at the corresponding eigenfrequency 51.15 GHz in Fig. 10. Here, we note that compared with the FSBS case, there is one more strain field $S_{xz}$ involved in BSBS. According to the symmetry properties of these elastic fields, as well as those of the electric fields, it can be checked that both the PE and MI AOCCs still vanish. The detailed analysis is similar to the one presented above. Here, to avoid redundancy, we do not go further on it.

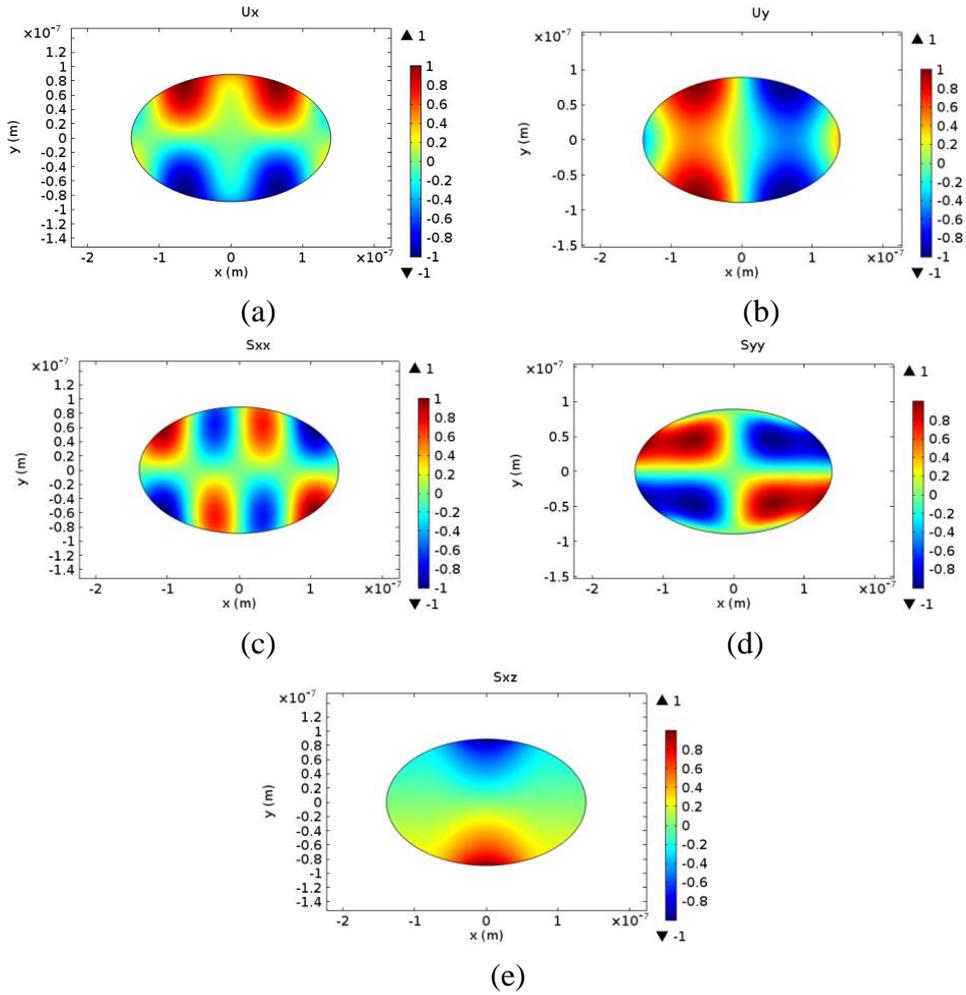

Fig. 10 Different field components (a) $U_x$, (b) $U_y$, (c) $S_{xx}$, (d) $S_{yy}$ and (e) $S_{xz}$ of the elastic mode with the eigenfrequency 51.15 GHz corresponding to the case of γ=0° in Fig. 4(b)

In fact, as reflected by Fig. 6, if the waveguide axes are aligned with the principal material axes (γ=0° or 90°), then the axisymmetric or anti-axisymmetric structures of the optical and elastic fields would result in possible vanishing of AOCCs. And additionally, if there is a small change of γ, the axial symmetries of the wave fields can be effectively broken, resulting in a dramatic increase of the magnitudes of

AOCCs and SBS gains. However, if the waveguide axes are far from being aligned with the principal material axes, then the wave fields do not have axial symmetries, resulting in much smaller possibilities for the AOCCs to vanish. Therefore, as reflected by Fig. 7, the kind of γ-sensitive resonant peaks can hardly be found.

## 5. Concluding remarks

In this work, the integral formulae for calculating the AOCCs due to the PE and MI effects in SBS are extended to an optically anisotropic waveguide. Then, based on the extended formulae, the SBSs in an elliptical nanowire made of the strongly birefringent material rutile are simulated. The extraordinary axis of rutile is assumed to be inside the waveguide's transverse plane, so that a strong transverse anisotropy is realized. The numerical results show that:
  (1) The SBS gains and the corresponding phononic frequencies are both significantly influenced by the orientation of the waveguide cross-section with respect to the principal material axes.
  (2) More interestingly, the SBS gains of some elastic modes are found to be very sensitive to the small misalignment between the waveguide axes and the principal material axes.

Detailed physical interpretations of this interesting phenomenon are provided. If the waveguide axes are aligned with the principal material axes, the axisymmetric or anti-axisymmetric structures of the optical and elastic fields are quite possible to result in vanishing AOCCs. And on the other hand, a small misalignment between the two kinds of axes can effectively break the axisymmetries or anti-axisymmetries of the fields, resulting in dramatic increase of the magnitudes of AOCCs and SBS gains.

This interesting phenomenon implies an attractive way for more sensitive tuning of the SBS gain without significantly changing the phononic frequency, which can be implemented through a non-circular sub-wavelength waveguide with strong transverse anisotropies. Therefore, material anisotropy may provide a new design freedom in some state of the art applications of the SBS at sub-wavelength scales.

## Acknowledgement

This work is supported by the National Natural Science Foundation of China (Project No. 11372031). The first author also acknowledges the support by the China Scholarship Council (CSC).

## Appendix A: detailed derivation for the MI acousto-optical coupling coefficient

Needless to say, Eq. (24) still holds if the optical field components appearing in it are replaced by their complex amplitudes. Then, we should have

$$\Delta \tilde{\mathbf{e}}_{\perp,\text{MI}} = \left( \tilde{e}_{y'}^{\text{I}} - \tilde{e}_{y'}^{\text{II}} \right) \hat{n}_{y'} \tag{A1a}$$

$$= \left[ \left( \zeta_{31}^{\text{I}} - \zeta_{31}^{\text{II}} \right) \tilde{e}_{\parallel,x'} + \left( \zeta_{32}^{\text{I}} - \zeta_{32}^{\text{II}} \right) \tilde{e}_{\parallel,z} + \left( \zeta_{33}^{\text{I}} - \zeta_{33}^{\text{II}} \right) \tilde{d}_{y'} \right] \hat{n}_{y'}, \tag{A1b}$$

$$\Delta \tilde{\mathbf{d}}_{\parallel,\text{MI}} = \left( \tilde{d}_{\parallel,x'}^{\text{I}} - \tilde{d}_{\parallel,x'}^{\text{II}} \right) \hat{n}_{\parallel,x'} + \left( \tilde{d}_{\parallel,z}^{\text{I}} - \tilde{d}_{\parallel,z}^{\text{II}} \right) \hat{n}_{\parallel,z} \tag{A2a}$$

$$= \left[ \left( \zeta_{11}^{\text{I}} - \zeta_{11}^{\text{II}} \right) \tilde{e}_{\parallel,x'} + \left( \zeta_{12}^{\text{I}} - \zeta_{12}^{\text{II}} \right) \tilde{e}_{\parallel,z} + \left( \zeta_{13}^{\text{I}} - \zeta_{13}^{\text{II}} \right) \tilde{d}_{y'} \right] \hat{n}_{\parallel,x'} +$$

$$\left[ \left( \zeta_{21}^{\text{I}} - \zeta_{21}^{\text{II}} \right) \tilde{e}_{\parallel,x'} + \left( \zeta_{22}^{\text{I}} - \zeta_{22}^{\text{II}} \right) \tilde{e}_{\parallel,z} + \left( \zeta_{23}^{\text{I}} - \zeta_{23}^{\text{II}} \right) \tilde{d}_{y'} \right] \hat{n}_{\parallel,z}, \tag{A2b}$$

So far, we have obtained Eqs. (A1b) and (A2b), i.e., the representations of the changes $\Delta \tilde{\mathbf{e}}_{\perp,\text{MI}}$ and $\Delta \tilde{\mathbf{d}}_{\parallel,\text{MI}}$ of the discontinuous optical field components in terms of the continuous optical field components $\tilde{\mathbf{e}}_{\parallel}$ and $\tilde{\mathbf{d}}_{\perp}$. Then, the substitution of them into the contour-integral formulae Eq. (14) for the MI AOCC yields

$$C_{\text{MI}} = \oint_{\Sigma} \left\{ \left[ \left( \zeta_{11}^{\text{I}} - \zeta_{11}^{\text{II}} \right) \tilde{e}_{\parallel,x'}^{(2)} + \left( \zeta_{12}^{\text{I}} - \zeta_{12}^{\text{II}} \right) \tilde{e}_{\parallel,z}^{(2)} + \left( \zeta_{13}^{\text{I}} - \zeta_{13}^{\text{II}} \right) \tilde{d}_{y'}^{(2)} \right] \left( \tilde{e}_{\parallel,x'}^{(1)} \right)^* + \right.$$

$$\left[ \left( \zeta_{21}^{\text{I}} - \zeta_{21}^{\text{II}} \right) \tilde{e}_{\parallel,x'}^{(2)} + \left( \zeta_{22}^{\text{I}} - \zeta_{22}^{\text{II}} \right) \tilde{e}_{\parallel,z}^{(2)} + \left( \zeta_{23}^{\text{I}} - \zeta_{23}^{\text{II}} \right) \tilde{d}_{y'}^{(2)} \right] \left( \tilde{e}_{\parallel,z}^{(1)} \right)^* -$$

$$\left. \left[ \left( \zeta_{31}^{\text{I}} - \zeta_{31}^{\text{II}} \right) \tilde{e}_{\parallel,x'}^{(2)} + \left( \zeta_{32}^{\text{I}} - \zeta_{32}^{\text{II}} \right) \tilde{e}_{\parallel,z}^{(2)} + \left( \zeta_{33}^{\text{I}} - \zeta_{33}^{\text{II}} \right) \tilde{d}_{y'}^{(2)} \right] \left( \tilde{d}_{y'}^{(1)} \right)^* \right\} \left[ (\tilde{\mathbf{u}})^* \cdot \hat{\mathbf{n}}_{y'} \right] dl. \tag{A3}$$

Based on the interesting properties of the matrix $\boldsymbol{\zeta}$ shown by Eq. (26), we further have

$$C_{\text{MI}} = \oint_{\Sigma} \left\{ \left( \zeta_{11}^{\text{I}} - \zeta_{11}^{\text{II}} \right) \left[ \tilde{e}_{\parallel,x'}^{(2)} \left( \tilde{e}_{\parallel,x'}^{(1)} \right)^* \right] + \left( \zeta_{22}^{\text{I}} - \zeta_{22}^{\text{II}} \right) \left[ \tilde{e}_{\parallel,z}^{(2)} \left( \tilde{e}_{\parallel,z}^{(1)} \right)^* \right] + \left( \zeta_{12}^{\text{I}} - \right. \right.$$

$$\left. \zeta_{12}^{\text{II}} \right) \left[ \tilde{e}_{\parallel,z}^{(2)} \left( \tilde{e}_{\parallel,x'}^{(1)} \right)^* + \tilde{e}_{\parallel,x'}^{(2)} \left( \tilde{e}_{\parallel,z}^{(1)} \right)^* \right] + \left( \zeta_{13}^{\text{I}} - \zeta_{13}^{\text{II}} \right) \left[ \tilde{e}_{\parallel,x'}^{(2)} \left( \tilde{d}_{y'}^{(1)} \right)^* + \left( \tilde{e}_{\parallel,x'}^{(1)} \right)^* \tilde{d}_{y'}^{(2)} \right] +$$

$$\left( \zeta_{23}^{\text{I}} - \zeta_{23}^{\text{II}} \right) \left[ \tilde{e}_{\parallel,z}^{(2)} \left( \tilde{d}_{y'}^{(1)} \right)^* + \tilde{d}_{y'}^{(2)} \left( \tilde{e}_{\parallel,z}^{(1)} \right)^* \right] - \left( \zeta_{33}^{\text{I}} - \zeta_{33}^{\text{II}} \right) \left[ \tilde{d}_{y'}^{(2)} \left( \tilde{d}_{y'}^{(1)} \right)^* \right] \right\} \left[ (\tilde{\mathbf{u}})^* \cdot \hat{\mathbf{n}}_{y'} \right] dl \tag{A4}$$

Then, substituting the elements of the matrix $\boldsymbol{\zeta}$ given in Eq. (25b) into Eq. (A4) eventually arrives at Eq. (27), i.e. the contour-integral formula for the AOCC due to the MI effect in an anisotropic waveguide.

## References

[1] R. W. Boyd, *Nonlinear Optics, 2$^{nd}$ ed.* (Academic Press, 2003).
[2] M. J. Damen, V. I. Vlad, V. Babin and A. Mocofanescu, *Stimulated Brillouin Scattering: Fundamentals and Applications* (CRC Press, 2003).

[3] M. F. Ferreira, *Nonlinear Effects in Optical Fibers*, (John Wiley & Sons Inc., 2011).
[4] X. Bao, "Optical Fiber Sensors Based on Brillouin Scattering", Opt. Phot. News 20, 40-45 (2009).
[5] A. Zadok, A. Eyal and M. Tur, "Stimulated Brillouin scattering slow light in optical fibers", App. Opt. 50, 38-49 (2011).
[6] V. P. Kalosha, L. Chen and X. Bao, "Slow and fast light via SBS in optical fibers for short pulses and broadband pump", Opt. Exp. 14, 12693 (2006).
[7] R. Shelby, M. Levenson and P. Bayer, "Guided acoustic-wave Brillouin scattering", Phys. Rev. B 31, 5244 (1985).
[8] P. Dainese, P. Russell, N. Joly, J. Knight, G. Wiederhecker, H. Fragnito, V. Laude, and A. Khelif, "Stimulated Brillouin Scattering from Multi-GHz-Guided Acoustic Phonons in Nanostructured Photonic Crystal Fibres", Nat. Phys. 2, 388-392 (2006).
[9] J. C. Beugnot and V. Laude, "Electrostriction and guidance of acoustic phonons in optical fibers", Phys. Rev. B 86, 224304 (2012).
[10] V. Laude and J. C. Beugnot, "Generation of phonons from electrostriction in small-core optical waveguides", AIP Advances 3, 042109 (2013).
[11] D. T. Hon, "Pulse compression by stimulated Brillouin scattering", Opt. Lett. 5, 516-518 (1980).
[12] I. Velchev, D. Neshev, W. Hogervorst and W. Ubachs, "Pulse compression to the subphonon lifetime region by half-cycle gain in transient stimulated Brillouin scattering", IEEE J. Quantum Electron. 35, 1812-1816 (1999).
[13] M. Eichenfield, J. Chan, R. Camacho, K.J. Vahala and O. Painter, "Optomechanical crystals", Nature 462, 78-82 (2009).
[14] M. Eichenfield, *Cavity optomechanics in photonic and phononic crystals: engineering the interaction of light and sound at the nanoscale* (Ph. D. Dissertation, California Institute of Technology, 2010).
[15] K. C. Balram, M. I. Davanço, J. D. Song and K. Srinivasan, "Coherent coupling between radiofrequency, optical and acoustic waves in piezo-optomechanical circuits", Nat. Photon. 10, 346-352 (2016).
[16] P. T. Rakich, C. Reinke, R. Camacho, P. Davids and Z. Wang, "Giant enhancement of stimulated Brillouin scattering in the subwavelength limit", Phys. Rev. X 2, 011008 (2012).
[17] H. Shin, W. Qiu, R. Jarecki, J. A. Cox, R. H. Olsson, A. Starbuck, Z. Wang and P. T. Rakich, "Tailorable stimulated Brillouin scattering in nanoscale silicon waveguides", Nat. Commun. 4, 1944 (2013).
[18] W. Qiu, P. T. Rakich, H. Shin, H. Dong, M. Soljačić and Z. Wang, "Stimulated Brillouin scattering in nanoscale silicon step-index waveguides: a general framework of selection rules and calculating SBS gain", Opt. Exp. 21, 31402 (2013).
[19] R. V. Laer, B. Kuyken, D. V. Thourhout and R. Baets, "Analysis of enhanced stimulated Brillouin scattering in silicon slot waveguides", Opt. Lett. 39, 1242-1245 (2014).
[20] C. Wolff, R. Soref, C. G. Poulton and B. J. Eggleton, "Germanium as a material for stimulated Brillouin scattering in the mid-infrared", Opt. Exp. 22, 30735 (2014).

[21] R. V. Laer, B. Kuyken, V. Thourhout and R. Baets, "Interaction between light and highly confined hypersound in a silicon photonic nanowire", Nat. Photonics 9, 199-203 (2015).

[22] S. Reza. Mirnaziry, C. Wolff, M. J. Steel, B. J. Eggleton and C. G. Poulton, "Stimulated Brillouin scattering in silicon/chalcogenide slot waveguides", Opt. Exp. 24, 4786 (2016)

[23] C. J. Sarabalis, J. T. Hill and A. H. Safavi-Naeini, "Guided acoustic and optical waves in silicon-on-insulator for Brillouin scattering and optomechanics", APL Photon. 1, 071301 (2016).

[24] G. D. Chen, R. W. Zhang, J. Q. Sun, H. Xie and Y. Gao, D. Q. Feng and H. Xiong, "Mode conversion based on forward stimulated Brillouin scattering in a hybrid phononic-photonic waveguide", Opt. Exp. 22, 32060 (2014).

[25] R. W. Zhang, G. D. Chen and J. Q. Sun, "Analysis of acousto-optic interaction based on forward stimulated Brillouin scattering in hybrid phononic-photonic waveguides", Opt. Exp. 24, 13051 (2016).

[26] O. Florez, P. F. Jarschel, Y. A. V. Espinel, C. M. B. Cordeiro, T. P. Mayer Alegre, G. S. Wiederhecker and P. Dainese, "Brillouin scattering self-cancellation", Nat. Commun. 7, 11759 (2016).

[27] J. C. Beugnot, S. Lebrun, G. Pauliat, H. Maillotte, V. Laude and T. Sylvestre, "Brillouin light scattering from surface acoustic waves in a subwavelength-diameter optical fibre", Nat. Commun. 5, 5242 (2014).

[28] C. G. Poulton, R. Pant and B. J. Eggleton, "Acoustic confinement and stimulated Brillouin scattering in integrated optical waveguides", J. Opt. Soc. Am. B 30, 2657 (2013).

[29] R. Pant, C. G. Poulton, D. Y. Choi, H. Mcfarlane, S. Hile, E. Li, L. Thevenaz, B. Luther-Davies, S. J. Madden and B. J. Eggleton, "On-chip stimulated Brillouin scattering", Opt. Exp. 25, 8285 (2011).

[30] C. G. Poulton, R. Pant, A. Byrnes, S. Fan, M. J. Steel and B. J. Eggleton, "Design for broadband on-chip isolator using stimulated Brillouin scattering in dispersion-engineered chalcogenide waveguides", Opt. Exp. 20, 21235 (2012).

[31] I. V. Kabakova, R. Pant, D. Y. Choi, S. Debbarma and B. Luther-Davies, Stephen J. Madden,2,3 and Benjamin J. Eggleton, "Narrow linewidth Brillouin laser based on chalcogenide photonic chip", Opt. Lett. 38, 3208-3211 (2013).

[32] B. J. Eggleton, C. G. Poulton and R. Pant, "Inducing and harnessing stimulated Brillouin scattering in photonic integrated circuits", Adv. Opt. Phot. 5, 536-587 (2013).

[33] V. Laude and J. C. Beugnot, "Lagrangian description of Brillouin scattering and electrostriction in nanoscale optical waveguides", New J. Phys. 17, 125003, (2015).

[34] C. Wolff, M. J. Steel, B. J. Eggleton and C. G. Poulton, "Acoustic build-up in on-chip stimulated Brillouin scattering", Sci. Rep. 5, 13656 (2015).

[35] C. Wolff, P. Gustsche, M. J. Steel, B. J. Eggleton and C. G. Poulton, "Impact of nonlinear loss on stimulated Brillouin scattering", J. Opt. Soc. Am. B 32, 1968-1978 (2015).

[36] C. Wolff, M. J. Steel, B. J. Eggleton and C. G. Poulton, "Stimulated Brillouin

scattering in integrated photonic waveguides: Forces, scattering mechanisms and coupled-mode analysis", Phys. Rev. A 92, 013836 (2015).
[37] J. E. Sipe and M. J. Steel, "A Hamiltonian treatment of stimulated Brillouin scattering in nanoscale integrated waveguides", New J. Phys. 18, 045004 (2016).
[38] C. Wolff, R. V. Laer, M. J. Steel, B. J. Eggleton and C. G. Poulton, "Brillouin resonance broadening due to structural variations in nanoscale waveguides", New J. Phys. 18, 025006 (2016).
[39] R. V. Laer, R. Baets and D. V. Thourhout, "Unifying Brillouin scattering and cavity optomechanics", Phys. Rev. A 93, 053828 (2016).
[40] K. P. Huy, J. C. Beugnot, J. C. Tchahame and T. Sylvestre, "Strong coupling between phonons and optical beating in backward Brillouin scattering", Phys. Rev. A 94, 043847 (2016).
[41] C. Wolff, B. Stiller, B. J. Eggleton, M. J. Steel and C. G. Poulton, "Cascaded forward Brillouin scattering to all Stokes orders", New. J. Phy. 19, 023021 (2017).
[42] M. J. A. Smith, C. Wolff, C. M. D. Sterke, M. Lapine, B. T. Kuhlmey and C. G. Poulton, "Stimulated Brillouin scattering in metamaterials", J. Opt. Soc. Am. B, 2162-2171 (2016).
[43] M. J. A. Smith, C. Wolff, C. M. D. Sterke, M. Lapine, B. T. Kuhlmey and C. G. Poulton, "Stimulated Brillouin scattering enhancement in silicon inverse opal waveguides", Opt. Exp. 24, 25148 (2016).
[44] Y. Pennec, V. Laude, N. Papanikolaou, B. Djafari-Rouhani, M. Oudich, S. E. Jallal, J. C. Beugnot, J. M. Escalante and A. Martńez, "Modeling light-sound interaction in nanoscale cavities and waveguides", Nanophotonics 3, 413-440, 2014.
[45] S. G. Johnson, M. Ibanescu, M. A. Skorobogatiy, O. Weisberg, J. D. Joannopoulos and Y. Fink, "Perturbation theory for Maxwell's equations with shifting material boundaries", Phys. Rev. E 65, 066611 (2002).
[46] B. Djafari-Rouhani, S. El-Jallal and Yan Pennec, "Phoxonic crystals and cavity optomechanics", Comp. Rend. Phys. 17, 555-564 (2016).
[47] T. X. Ma, K. Zou, Y. S. Wang, Ch. Zhang, and X. X. Su, "Acousto-optical interaction of surface acoustic and optical waves in a two-dimensional phoxonic crystal hetero-structure cavity", Opt. Exp. 23, 28443 (2014).
[48] T. X. Ma, Y. S. Wang and Ch. Zhang, "Enhancement of acousto-optical coupling in two-dimensional air-slot phoxonic crystal cavities by utilizing surface acoustic waves", Phys. Lett. A 381, 323-329 (2017).
[49] P. T. Rakich, Z. Wang, and P. Davids, "Scaling of Optical Forces in Dielectric Waveguides: Rigorous Connection between Radiation Pressure and Dispersion", Opt. Lett. 36, 217-219 (2011).
[50] P. T. Rakich, P. Davids, and Z. Wang, "Tailoring Optical Forces in Waveguides through Radiation Pressure and Electrostrictive Forces", Opt. Exp. 18, 14439 (2010).
[51] A. G. Krause, J. T. Hill, M. Ludwig, A. H. Safavi-Naeini, J. Chan, F. Marquardt and O. Painter, "Nonlinear Radiation Pressure Dynamics in an Optomechanical Crystal", Phys. Rev. Lett. 115, 233601 (2015).
[52] A. Yariv and P. Yeh, *Optical waves in crystals: Propagation and control of laser radiation* (John Wiley and Sons, 1984)

[53] Z. F. Bi, L. Wang, X. H. Liu, S. M. Zhang, M. M. Dong, Q. Z. Zhao, X. L. Wu and K. M. Wang, "Optical waveguides in TiO2 formed by He ion implantation", Opt. Exp. 20, 6712 (2012).

[54] C. C. Evans, *Nonlinear optics in titanium dioxide: from bulk to integrated optical devices* (Ph. D. Dissertation, Harvard University, 2013).

[55] C. C. Evans, C. Liu, and J. Suntivich, "Low-loss titanium dioxide waveguides and resonators using a dielectric lift-off fabrication process", Opt. Exp. 23, 11160 (2015).

[56] The material parameters used for rutile are listed as follows. The optical refractive indices $n_o$=2.494 and $n_e$=2.76; mass density $\rho$=2170 kg/m$^3$; elastic constants (in units of GPa): $C_{11}$=270, $C_{12}$=147, $C_{13}$=176, $C_{22}$=480, $C_{44}$=$C_{66}$=124 and $C_{55}$=193; and photoelastic constants: $p_{11}$=-0.011, $p_{12}$=-0.168, $p_{13}$=0.172, $p_{21}$=-0.0965, $p_{22}$=-0.058, $p_{44}$=$p_{66}$=0 and $p_{55}$=0.072. Note that according to the definitions of the geometric axes shown in Fig. 2, the subscripts "1" and "3" of the material property tensors are interchangeable.

[57] http://www.korth.de/index.php/162/items/34.html